\documentclass[aps,prl,twocolumn,groupedaddress,showpacs]{revtex4}
\usepackage{graphicx}
\usepackage{epstopdf}
\usepackage{amssymb, amsmath}
\usepackage{multirow}
\usepackage{dcolumn}
\usepackage{bm}
\usepackage{subfigure}
\usepackage{color}
\usepackage{ulem}

\def\imagetop#1{\vtop{\null\hbox{#1}}}

\begin{document}
 
\title{Quantum Monte Carlo calculations in solids with downfolded Hamiltonians}
 
\author{Fengjie Ma, Wirawan Purwanto, Shiwei Zhang, and Henry Krakauer}
\affiliation{Department of Physics, College of William and Mary, Williamsburg, VA 23187}

\begin{abstract}

We present a systematic downfolding many-body approach for extended systems.  
Many-body calculations operate on a simpler Hamiltonian 
which retains material-specific properties. 
The Hamiltonian is systematically improvable and allows one to dial, in principle, between the simplest model and the original Hamiltonian. 
As a by-product, 
pseudopotential errors are essentially eliminated using a frozen-core treatment.
The computational cost of the many-body calculation is dramatically reduced without sacrificing accuracy.
We use the auxiliary-field quantum Monte Carlo (AFQMC) method to solve the downfolded Hamiltonian.
Excellent accuracy is achieved for a range of solids, including semiconductors, ionic insulators, and metals.
We further test the method by determining the spin gap in NiO, a challenging prototypical material with strong
electron correlation effects. 
This approach greatly extends the reach of general, \textit{ab initio} many-body calculations in  
materials. 

\end{abstract}


\maketitle

Developing accurate and efficient 
computational approaches for quantum matter 
has been a long-standing challenge. 
Parameter-free, material-specific many-body calculations are needed  where 
simpler methods, such as those based on density functional theory (DFT) \cite{KohnNobel} or perturbative approaches, break down. 
Examples range from strongly correlated
materials, such as transition metal oxides, to 
bond-stretching or
bond-breaking in otherwise moderately correlated systems. 
Quantum Monte Carlo (QMC) has become increasingly important in this regard \cite{QMCRMP,LectureNotes2013,AFQMCbond,AFQMCCr2,LukeSolids,DMCConds,DMCTMOs,DMCCuprates,AFQMCExcited}.  
However, systematic and routine applications of QMC in realistic 
materials still face major challenges. 
Here we present an approach which 
overcomes several of the obstacles and  
advances the capabilities of non-perturbative
ground-state calculations in correlated materials in general.

Our approach treats
downfolded Hamiltonians expressed with respect to a truncated basis set of 
mean-field orbitals of the
target system, using an auxiliary-field quantum Monte Carlo (AFQMC) method \cite{PhaselessAFQMC,AFQMCgaussian,LectureNotes2013}.
This allows QMC calculations to be performed with a much simpler Hamiltonian 
while retaining material-specific properties. 
The simplification, often with drastic reduction in computational cost, 
can extend the reach of \textit{ab initio} computations to 
more complex materials.
A large gain in statistical accuracy often results as well, 
because of the smaller range 
of energy scales (or many fewer degrees of freedom) which need to be sampled stochastically
in the downfolded Hamiltonian. 

Two other key advantages follow as a result of this approach. 
First, by varying the cut-off that controls the truncation of the basis orbitals, one
could in principle dial 
between the original full-basis Hamiltonian and the simplest model. QMC calculations can be performed at each 
stage. 
This allows a systematically 
improvable set of calculations that connect simple models to full materials specificity.
Second, the approach introduces a new way for treating core electrons, which has been a critical issue  in
QMC. 
Significant errors are often present with the use of pseudopotential (PSP) in QMC \cite{SorellaSi,SiDMC}, 
due to i) inherent limitations in the accuracy of such 
PSPs (single-projector, generated in an atomic environment, from independent-electron calculations)
\cite{TNDFpsp,BFDpsp,PWAFQMC}, and ii) approximations in how the PSP has to be implemented in standard 
diffusion Monte Carlo (DMC) calculations \cite{QMCRMP,DMCLocality}.
In a recent DMC study of high-pressure BN \cite{EslerDMC2010}, 
the reliable determination of
the equation of state (EOS) 
required all-electron calculations, which in 
most materials  would not be practical. 
In our approach a frozen-core (FC) treatment \cite{DownfoldingFC} is used to essentially eliminate
PSP errors. 

The most fundamental issue in computations of electron correlation effects is accuracy. 
For QMC calculations,
the fermion sign problem must, in all but a few special cases, be controlled with an approximation. 
The AFQMC framework, by carrying out the 
random walks in non-orthogonal Slater determinant space, has shown to lead to an approximation 
which is more accurate and less dependent on the trial wave function
\cite{LectureNotes2013,AFQMCmolecule,PWAFQMC,AFQMCExcited,AFQMCbtin,AFQMCbond,AFQMCgaussian}. 
In both lattice models \cite{Hub-SDW} and molecular systems \cite{AFQMCCr2}, 
recent advances with AFQMC have allowed systematic accuracy in treating strongly correlated systems.
The method proposed here provides an approach to seamlessly integrate these advances  
in the study of solids.
We illustrate the approach by obtaining accurate equilibrium properties
in a range of solids, including semiconductors, ionic insulators, and metals. 
We then show that  
the present approach can describe BN, with an accurate EOS extending to high pressures,
without resorting to all-electron 
calculations. Finally, the spin gap in 
strongly correlated NiO is accurately determined and compared with experiment.

\begin{figure}
    \begin{tabular}{lllc}
     \imagetop{(a)} & \imagetop{\includegraphics[width=0.14\textwidth]{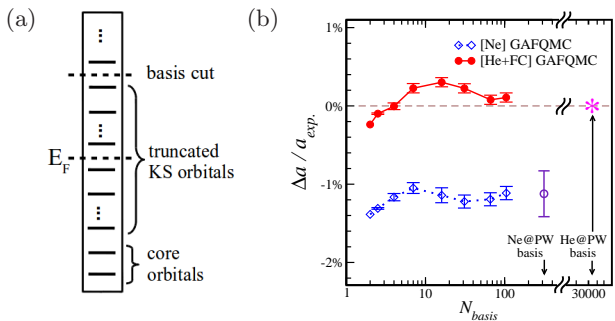}} & \imagetop{(b)} & \imagetop{\includegraphics[width=0.24\textwidth]{Si_Elas_v7.eps}} 
    \end{tabular}    
    
\caption{\label{trunbasis} (Color online)
a) Illustration for basis downfolding. Solid black lines represent DFT KS orbitals. 
A compact basis is constructed with DFT-KS orbitals, by neglecting the less  physically relevant 
high-energy states above a truncation energy.
Deep core electrons can be frozen at the mean-field 
level by a frozen-core treatment. 
b) Error in the 
calculated lattice constant 
in Si vs.~basis size. 
Results with standard Ne-core and a highly accurate He-core PSP 
plus FC are both shown. 
For the Ne-core PSP, the full planewave AFQMC result is indicated 
by the indigo open circle. For the He-core PSP, the number of planewaves required in the full calculation 
is indicated (note logarithmic scale).}
\end{figure}

The construction of the downfolded Hamiltonian begins with 
a standard DFT calculation for the target system. 
This is done using a planewave basis with PSPs.
(Extremely hard PSPs, e.g., He-core for third row elements or ``zero-electron-core" for the first-row, are employed, if necessary, to eliminate transferability errors of conventional norm-conserving PSPs~\cite{DownfoldingFC}, 
at essentially no additional computational cost in the ensuing many-body calculations.) Planewaves are desirable at this 
stage, because they provide an unbiased representation of 
the many-body Hamiltonian.
We then use
the Kohn-Sham (KS) orbitals as basis set, tuned
to eliminate less physically relevant high-energy virtual states and low-energy core states,
as illustrated in Fig.~\ref{trunbasis}a.
Expressed in this basis, the effective downfolded Hamiltonian is given by one-body and two-body terms
whose matrix elements are
\begin{equation}
K_{ij}=\langle \chi_i| \hat{K} |\chi_j\rangle; \qquad
V_{ijkl}=\langle \chi_i \chi_j |{\hat V}| \chi_k \chi_l\rangle
\label{eq:MEs}
\end{equation}
where $|\chi_i\rangle$ is a KS orbital, the labels $i$, $j$, $k$, and $l$ all run in the truncated basis set,
$\hat K$ includes all 
one-body (and constant) terms in the Hamiltonian 
and $\hat V$ is the two-body interaction term. The matrix elements, 
which encode the periodicity and 
the Coulomb interaction in the underlying supercell \cite{PWAFQMC},
can be conveniently computed using
fast Fourier transforms, as the orbitals $\chi$ are given in planewaves. 
We use twist boundary conditions \cite{QMCTwist} on the supercell. 
For inversion-symmetric systems, all the matrix elements are \textit{real} under 
any twist ${\mathbf k}$. 
The core 
states can be frozen in the corresponding  
KS orbitals of the solid;
two-body core-valence interactions appear as one-body ion-valence terms $K_{ij}$
and core-only interactions (one-body and two-body) contribute a constant \cite{DownfoldingFC}.

The downfolded Hamiltonian defined in Eq.~(\ref{eq:MEs}) is then treated using phaseless AFQMC 
\cite{PhaselessAFQMC} but in the molecular formalism \cite{AFQMCgaussian,Purwanto2011},
which can handle any one-particle basis functions.
The approach is illustrated for fcc Si in Fig.~\ref{trunbasis}(b), which shows the 
convergence of the calculated equilibrium lattice constant. 
Results are shown for both Ne- and He-core PSPs~\cite{opium}. 
With the Ne-core PSP, each Si atom contributes four electrons, 
and there are
no ``core electrons" in the diagram in Fig.~\ref{trunbasis}a. 
The basis cut controls the number of KS orbitals in the truncated basis, $N_{\rm basis}$.
When all states are retained in the truncation, the 
KS orbital basis is just a unitary transformation of the original planewave basis. 
As $N_{\rm basis}$ is increased, the result converges to the full planewave AFQMC 
result. The statistical error bar with the downfolded Hamiltonian is much smaller, however,
because many fewer auxiliary-fields need to be sampled 
\cite{PWAFQMC,Purwanto2011}. 
With the 
He-core PSP, very small radial cut-offs ($0.54$, $0.68$, and $0.54$~bohr for 
$s$, $p$, and $d$ channels, respectively)
were used, 
which resulted in a large planewave cut-off 
\mbox{$E_\mathrm{cut}=600$\,Ry}. 
The $2s$ and $2p$ electrons are then treated as ``core electrons,"
frozen in their KS orbitals. 
As seen in Fig.~\ref{trunbasis}(b), 
this approach (He-core plus FC) eliminates the $1.2\%$ error in the calculated 
lattice constant from the Ne-core PSP. Furthermore, the calculation reaches convergence with $\sim 100$ basis 
functions, more than two orders of magnitude smaller than 
would be required in the full planewave calculation.

\begin{figure}
\begin{center}
\includegraphics[width=0.35\textwidth]{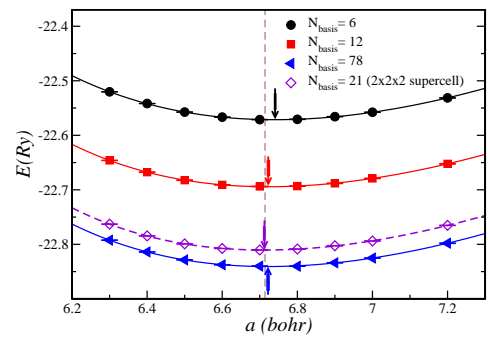}
\caption{\label{Diamond} (Color online) EOS of fcc diamond with different basis cut-offs, $N_{\rm basis}$ 
(per atom). 
The experimental lattice parameter is indicated by the vertical dashed 
line. The calculated equilibrium lattice constant for each curve 
is indicated by a solid arrow, with the width of the line showing 
the statistical error bar.
The total energy changes with $N_{\rm basis}$, but the equilibrium properties converge 
rapidly. 
Residual finite size effects are corrected and checked by calculations in
the larger supercell.
}
\end{center}
\end{figure}

The basis choice and truncation method are not unique. 
Possible truncation choices include a fixed number of basis functions, a fixed cut-off energy, a fixed ratio to the full basis, etc. 
We find that the 
first choice leads to the most rapid convergence in our EOS calculations.
There is also considerable freedom in the choice of underlying basis. 
In spin-polarized systems, we generate a spin-consistent basis set by 
diagonalizing the $2N_{\rm basis}\times 2N_{\rm basis}$ overlap matrix formed by 
$\langle \chi^\sigma_i|\chi^{\sigma'}_j\rangle$, where $\sigma$ and $\sigma'$ are spin indices. 
The resulting eigenfunctions corresponding to the largest $N_{\rm basis}$ eigenvalues are used as new ``KS orbitals," 
which leads
to an unbiased basis set and much faster convergence, as illustrated for NiO below.
Similarly, localization strategies could be applied to generate more efficient basis sets \cite{Hoyvik2013,NatOrbs,MLWOs}.

EOS calculations are shown in Fig.~\ref{Diamond} for fcc diamond. 
The total energy decreases with increasing $N_{\rm basis}$,
as expected, but the overall shape of the EOS is similar.
With each, a fit to the Murnaghan equation \cite{MurnaghanEOS} is done to obtain the 
equilibrium lattice constant and bulk modulus. 
With only \mbox{$N_\mathrm{basis}=12$} per atom, the equilibrium lattice constant is essentially converged. 
The calculations used ${\mathbf k}=\mbox{(0.5, 0.5, 0.5)}$;
one-body and two-body 
finite-size corrections derived from the corresponding simulation cells \cite{KZKFS,KZKFSspin}
are applied to the many-body results.  
As illustrated in the figure, calculations for a larger supercell were carried out to check 
that residual finite-size effects are smaller than the statistical errors in the final result.

\begin{figure}
\begin{center}
\includegraphics[width=0.38\textwidth, height=0.275\textwidth]{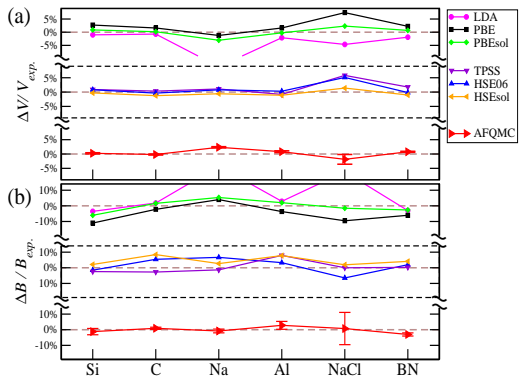}
\caption{\label{Tab1_fig_v0} 
Summary of calculated equilibrium volumes (a) 
and bulk moduli (b), shown as relative errors 
from experiment. Selected DFT results are also shown for reference.
Zero-point effects have been subtracted from the experiments~\cite{HSEsolZPAE,Datchi2007}.
}
\end{center}
\end{figure}

As additional tests in ``conventional'' systems, we 
apply the AFQMC downfolding approach to two metals Na and Al, an ionic crystal NaCl, and BN,
whose high-pressure EOS is further studied below 
for pressure calibration. 
Figure~\ref{Tab1_fig_v0} summarizes all the results of calculated equilibrium properties 
and compares them with experiment.
The calculations for NaCl and BN were analogous to those for Si and C.
The calculations in 
bcc Na and fcc Al used twist averaging over 90 random ${\mathbf k}$-points \cite{QMCTwist}. 
He-core PSPs were used for Na and Al, together with the FC treatment
as described for Si. 
This makes a major difference in both NaCl and Na. With a Ne-core PSP, the 
equilibrium volume is underestimated by $\sim 30\%$ in Na, for example. 
The error is eliminated by the FC approach, which allows the semi-core $2s$ and $2p$ electrons to fully relax in the \textit{target environment
of the solid} at the DFT-level, before freezing them in the corresponding KS orbitals in the many-body calculation.

The agreement between downfolding AFQMC and experiments \cite{TPSS,TPSSErr,TPSSZPAE,HSEsolZPAE}
is excellent. 
For reference, some representative DFT results are shown 
in Fig.~\ref{Tab1_fig_v0}:
the top panels in (a) and (b) include results from the widely used local-density (LDA) 
and generalized gradient (GGA, with two flavors, PBE and a variant which is specially designed for solids
and surfaces, PBEsol) 
approximations \cite{PZ,PBE,PBEsol};
the middle panels 
sample more recent developments in DFT, with
a meta-GGA (TPSS) and two flavors of hybrid functionals
(HSE06 and HSEsol) \cite{TPSS,TPSSErr,TPSSZPAE,HSEsolZPAE},
which are highly accurate in many conventional
systems but often involve empirical parameters. 
The AFQMC results (bottom panels) demonstrate that the new approach provides an 
\textit{ab initio},  parameter-free, many-body framework 
that is consistently accurate. 
The calculations used single-determinant trial wave functions taken directly from LDA or GGA
to control the sign or phase problem of the random walks in Slater determinant 
space \cite{PhaselessAFQMC,LectureNotes2013}. 
The systematic error from this approximation, based on extensive prior benchmarks \cite{PhaselessAFQMC,PWAFQMC,AFQMCExcited,AFQMCbtin}, is expected to be essentially negligible in these systems, in accord with the results in the figure. 
The largest uncertainty arises in NaCl and is statistical in nature. 
Different from the other systems, the ionic 
character results in valence states localized on the Cl atom. The high-energy virtual
KS orbitals, which are used to capture the effect of electron interactions, are free-electron like, however. As a result, convergence
of the EOS is slow and an extrapolation with respect to $1/N_\mathrm{basis}$ was needed to 
reach the complete basis set limit, 
resulting in larger 
uncertainty.
Clearly, this can be improved by using Wannier or other localized orbitals in the 
downfolding.

\begin{figure}
\begin{center}
\includegraphics[width=0.33\textwidth]{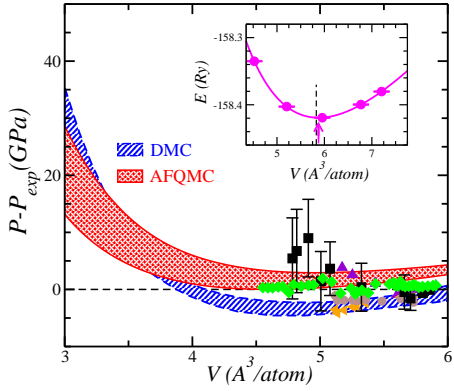}
\caption{\label{BN_EOS} (Color online) 
Pressure calibration and EOS in cubic BN. 
The main graph displays the calculated 
pressure vs.~volume at room-temperature, using the 
fitted experimental curve of Ref.~\onlinecite{Datchi2007} (green diamond symbols) as a 
pressure reference. 
Different symbols are from different experiments \cite{Datchi2007,Goncharov2007,Solozhenko1998,Knittle1989}.
The shading gives the overall statistical uncertainties in the calculations.
The all-electron 
DMC results are from Ref.~\onlinecite{EslerDMC2010}.
The inset shows the $T=0$\,K EOS  
near equilibrium from AFQMC.
The calculated equilibrium position is shown by the arrow.
The vertical 
line indicates the experimental value~\cite{Datchi2007,Goncharov2007}. 
}
\end{center}
\end{figure}

In a more demanding test, 
we apply downfolding AFQMC to obtain the EOS
of cubic BN for pressures up to 900\,GPa 
($V\sim 0.5 \,V_{eq}$).
This system has been identified as a promising material for an ultra-high pressure calibration scale \cite{EslerDMC2010,cBNnano}.
A recent DMC study stressed 
the need for all-electron (AE) calculations in order to 
obtain reliable results at high pressures \cite{EslerDMC2010}. 
The difficulty underscores the PSP transferability problem discussed above in the context of Na and Si,
and is exacerbated by the need to apply a locality approximation in DMC to treat non-local PSPs \cite{QMCRMP,DMCLocality}. 
The AE treatment would be difficult to realize for heavier atoms. 
Our calculations freeze the $1s$ electrons in their KS orbitals in the supercell at each volume, 
using extremely hard ``zero-electron-core" PSPs for B and N in the downfolding procedure \cite{BNPsp}. 
In most cases $\sim 55$ states/atom were used, but larger $N_{\rm basis}$ 
calculations were done at selected volumes to extrapolate the EOS to the complete basis set limit. 
We applied finite-temperature corrections following 
Ref.~\onlinecite{EslerDMC2010}. 
The calculations were done with ${\mathbf k}= \mbox{(0.5, 0.5, 0.0)}$ for 8-  and 16-atom supercells, 
with one- and two-body finite-size  corrections as discussed earlier; we have confirmed that residual errors 
are negligible compared to the final estimated error band, especially in the high-pressure regime. 
As seen in Fig.~\ref{BN_EOS}, the calculated EOS at low pressures 
is in excellent agreement with experiments 
~\cite{Datchi2007,Goncharov2007,Solozhenko1998,Knittle1989}.
The calculated equilibrium lattice constant, $6.820(3)$~bohr, is consistent with experimental measurements of $6.802$~bohr 
(zero-point energy removed), as shown in the inset. 
The EOS at low pressures shows small 
but discernible discrepancies with DMC results. 
Possible origins include differences in the finite-temperature corrections \cite{FTnote}, or
DMC fixed-node errors, 
and will require further investigation. 
At high pressures, the two QMC results are in good agreement, providing a consistent 
\textit{ab initio} pressure calibration.

\begin{figure}
\begin{center}
\includegraphics[width=0.38\textwidth]{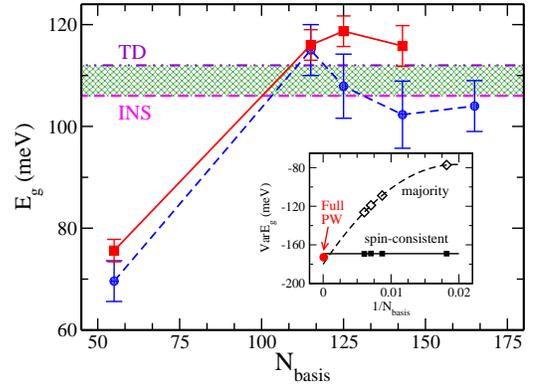}
\caption{\label{NiO_spingap} Spin gap in NiO, and comparison with experiment.
A smaller set of calculations (blue 
circles), averaging over two ${\mathbf k}$-points, confirms convergence 
with respect to basis set cutoff $N_{\rm basis}$. 
Calculations averaging over a $\mbox{$4 \times 4 \times 4$}$ Monkhorst-Pack grid (red squares)
are used to obtain the final results. 
The shaded bar represents the experimental range:  inelastic neutron scattering (INS) \cite{NiOINS} (bottom line)
and thermodynamic (TD)  measurements \cite{NiOTD} (top line).
The inset illustrates the much faster convergence enabled by the spin-consistent basis set (see text) than 
using the KS orbitals from the majority spin.  
}
\end{center}
\end{figure}

As a final application, we determine the spin gap
between the ferromagnetic (FM) state and the antiferromagnetic (AFM-II)  ground state in NiO.
Understanding and predicting magnetic properties of transition-metal oxides 
epitomizes the challenge of computations in quantum matter. 
NiO is a prototypical system for strong electron correlations. 
Many-body calculations of the spin gap have been limited, and DFT-based methods have yielded widely varying values \cite{NiODFT}. 
We use Ne-core and He-core PSPs for Ni and O, respectively. 
The downfolded Hamiltonian treats the Ni  $3s, \, 3p, \, 3d, \, 4s$ and O $2s, \, 2p$ electrons.
A rhombohedral supercell with a lattice constant of 4.17~$\AA$ containing two formula units is used.
To reduce one-body finite-size effects, 
we used twist-averaging with a $\mbox{$4 \times 4 \times 4$}$ Monkhorst-Pack grid \cite{MPgrid}.
(A recent study \cite{FCIQMC-nature} 
by full configuration-interaction QMC and coupled-cluster methods,
with calculations at ${\mathbf k}=\Gamma$, obtained a 
gap value $\sim 0.96\,eV$.) 
One- and two-body finite-size corrections \cite{KZKFSspin} are then applied to the many-body results
(two-body finite-size effects are greatly reduced by cancellation, because the two 
phases share the same supercell).
In the downfolding
we use the spin-consistent basis sets discussed earlier. 
In the inset in Fig.~\ref{NiO_spingap},
the variational gap value 
from the single-determinant trial wave functions 
is shown 
vs.~the number of basis functions, for both the spin-consistent basis set and 
one which uses truncated KS orbitals of the majority spin. 
Both converge to the same infinite basis-set limit, as expected, but the former 
greatly accelerates convergence. 
Note that the variational gap (which has been averaged over ${\mathbf k}$-points) is actually 
negative, 
i.e., the trial wave functions identify the incorrect phase for the ground state. The AFQMC calculations 
correctly recovers from these, and yield a final estimate of the gap of $116(3)$\,meV,
in good agreement with experiments \cite{NiOINS,NiOTD}.

In summary, we have presented a systematic downfolding Hamiltonian 
approach for solids. 
As a first test, parameter-free calculations of equilibrium properties are demonstrated 
in semiconductors, metals, and ionic insulators. 
QMC PSP errors are eliminated
without (prohibitive) all-electron calculations, as demonstrated in BN.
The spin gap in strongly correlated NiO is accurately determined.
The approach drastically reduces complexity and computational cost, and greatly
extends the reach of \textit{ab initio}, non-perturbative, many-body computations 
in complex materials.
Furthermore, the framework 
provides 
a 
tunable connection between the full materials-specific Hamiltonian and simplified models.
The downfolding approach can be generalized to carry out excited state and many-body band structure calculations, which was recently
formulated \cite{AFQMCExcited} in planewave AFQMC.
A large number of applications are possible within the present form. 
Further improvements, for example by using localized virtual states or optimizing the orbitals
with respect to the environments, will lead to even more general and powerful approaches.

This work is supported by DOE (DE-SC0008627 and DE-SC0001303), NSF (DMR-1409510), and ONR (N000141211042). 
An award of computer time was provided by the Innovative and Novel Computational Impact on Theory and Experiment (INCITE) program, using resources of the Oak Ridge Leadership Computing Facility at the Oak Ridge National Laboratory, which is supported by the Office of Science of the U.S. Department of Energy under Contract No. DE-AC05-00OR22725. We also acknowledge computing support from 
the computational facilities at the College of William and Mary.

\bibliography{gafqmc}

\begin{thebibliography}{51}
\bibitem[{BNP()}]{BNPsp}
\bibinfo{note}{The radial cut-offs are 0.58~bohr for $s$ and $p$ channels of B,
  and 0.61~bohr for those of N, resulting in planewave cut-off 600~Ry.}
\end{thebibliography}

\end{document}